\documentclass[graybox]{svmult}

\usepackage{mathptmx}       % selects Times Roman as basic font
\usepackage{helvet}         % selects Helvetica as sans-serif font
\usepackage{courier}        % selects Courier as typewriter font
\usepackage{type1cm}        % activate if the above 3 fonts are
\usepackage{makeidx}         % allows index generation
\usepackage{graphicx}        % standard LaTeX graphics tool
\usepackage{multicol}        % used for the two-column index
\usepackage[bottom]{footmisc}% places footnotes at page bottom

\makeindex             % used for the subject index
\begin{document}

\title*{The Art of Data Science}
\author{Matthew J. Graham}
\institute{Matthew J. Graham \at California Institute of Technology, 1200 E California Blvd, Pasadena, CA 91125, USA, \email{mjg@caltech.edu}}
\maketitle

\abstract*{Each chapter should be preceded by an abstract (10--15 lines long) that summarizes the content. The abstract will appear \textit{online} at \url{www.SpringerLink.com} and be available with unrestricted access. This allows unregistered users to read the abstract as a teaser for the complete chapter. As a general rule the abstracts will not appear in the printed version of your book unless it is the style of your particular book or that of the series to which your book belongs.
Please use the 'starred' version of the new Springer \texttt{abstract} command for typesetting the text of the online abstracts (cf. source file of this chapter template \texttt{abstract}) and include them with the source files of your manuscript. Use the plain \texttt{abstract} command if the abstract is also to appear in the printed version of the book.}

\abstract{To flourish in the new data-intensive environment of 21$^{st}$ century science, we need to evolve new skills. These can be expressed in terms of the systemized framework that formed the basis of mediaeval education -- the {\em trivium} (logic, grammar, and rhetoric) and {\em quadrivium} (arithmetic, geometry, music, and astronomy). However, rather than focusing on number, data is the new keystone. We need to understand what rules it obeys, how it is symbolized and communicated and what its relationship to physical space and time is. In this paper, we will review this understanding in terms of the technologies and processes that it requires. We contend that, at least, an appreciation of all these aspects is crucial to enable us to extract scientific information and knowledge from the data sets which threaten to engulf and overwhelm us.  }

\section{Introduction}
\label{sec:1}
Teaching in the great universities of the Middle Ages focussed on the seven liberal arts: the {\it trivium} - logic, grammar, and rhetoric -, and the {\it quadrivium} - arithmetic, geometry, music, and astronomy. Training and competency in these subjects was believed sufficient to form an individual with the necessary intellectual capabilities to pursue a career or further study in law, theology, or natural philosophy. Today's natural philosophers are schooled in the arts of empirical, theoretical, and computational scientific methodology as preparation for their professional careers. However, the vanguard of the data revolution is now upon us with high-dimensional, high-volume, feature-rich data sets becoming an increasingly common aspect of our everyday workplace and we are ill-prepared. 

To meet this challenge, a fourth paradigm \cite{fp} is emerging: the so-called data-intensive science or $x$-informatics (where $x$ is the science of choice, such as {\it bio}-informatics, {\it geo}informatics or {\it astro}informatics), which will support and drive scientific discovery in the 21$^{st}$ century. This is not just an incremental development on what has gone before but something entirely new and we're still trying to figure out not only what shape it is and where its boundaries lie but, more fundamentally, what its basic rules are. Yet, at the same time, it would not be unfamiliar to a 13$^{th}$ century scholar.

The core of the mediaeval syllabus was a systemization of knowledge - what rules does it obey, how is it symbolized and how is it communicated - and, in particular, numerical knowledge and the relationship of number to physical space and time. Arithmetic, for example, was the study of pure number whereas music was the study of number in relation to time \cite{kline}. In this paper, we aim to show how the new art of data science can similarly be framed as a systemization of {\it data} and its relationship to space and time, particularly in regard to its technological aspects. Those this has relevancy to many sciences, our broad theme will be astronomy.

\section{The logic of data} % What rules does it obey?
\label{sec:2}
Just as alchemists thought of mercury as the {\it prima materia} (first matter) from which all metals were formed, so scientists consider data to be the basis of all understanding. Yet it is a commodity as fluid and elusive as its elemental counterpart. Great cost and effort is expended by empiricists to measure it, computationalists to imitate it and theoreticians to formulate it but, even then, do we really understand what we are working with? Even the word itself is open to speculation \cite{norman}. The legitimate use and, especially reuse, of data requires context: not just the raw processing of numerical or symbolic values but also adequate attention to their origins, systematics, biases, and bounds. 

Hogg \& Lang \cite{hogglang1, hogglang2} argue that most of astronomy has been conducted through catalogues, an inferior data product, derived from raw data but missing the necessary knowledge about the data -- how it was analysed, errors estimated, etc. -- to support any sophisticated statistical inferencing, such as resolving deblending issues in SDSS. Anything beyond raw data values is metadata and needs to be sufficiently described, preferably in terms of a (Bayesian) posterior probability model, so that arbitrary questions (cast as hypotheses) can be asked of it with maximal usage of the available information. Taken to its extreme, the ultimate model would be of the entire sky through wavelength and time from which any astronomical image ever taken at any time with any equipment in any configuration could be generated and thus anomalies in any data easily identified.

Semantics provides an alternative but complementary approach, framing knowledge about data in terms of programmable structures rather than likelihood functions. Semantic constructs such as ontologies allow domain knowledge to be expressed in a machine-processible format in terms of classes, properties (relationships) and operations and data as instances of these. Logical inferencing over the classes and the instances allows inconsistencies in both the data and its description to be determined. Semantics also aids reusing data: different descriptions/interpretations can be efficiently combined/reconciled {\it by machines} to construct data sets of wider scope and applicability than originally intended. For example, the study of spectral energy distributions using 
multiwavelength data sets formed by combining (heterogeneous) single filter/passband observations of astronomical objects needs a proper treatment of flux values in each component data set, which information would be encoded in their ontologies or equivalent structures. 

We should never just blindly work with data, particularly as it becomes ever more complex. Explorations may skirt around full and proper treatments but understanding the rules that data obeys and manipulating this logic through inferencing, be it statistical or logical, is necessary for validatable and replicable discovery. Developing such systems and ensuring that they are performant in the face of forthcoming data expectations is a real challenge, however, and not one to be easily glossed over but it is one that can be met by interdisciplinary engagement across the vertical silos of individual sciences.

\section{The grammar of data} % How is it symbolized?
\label{sec:3}
To the mediaeval mind, unravelling the mysteries of the world lay in decoding the symbolic languages that Nature employed to hide her secrets. Everything was charged with meaning, be it through number, colour, geometry, or some more subtle aspect or property. The wise man could read the hidden messages (the patterns in a monastery garden) whereas the fool saw just the forms (the flowers), understanding nothing further of their meaning. The symbolism of data is far more profane: complex objects are converted to sequences of bits for persistence and communication but there are still a variety of representations (data serialization formats), each with a specific meaning and purpose. 

At its base level, data is comprised of numbers or symbols, normally stored in a digital (binary) representation. Whilst every piece of data could just be treated as an amorphous chunk of bits, the utility of this approach is really limited to large data objects (blobs), such as the streaming multimedia that forms an increasing fraction of web traffic. Data is far more manipulable if it is structured in some way and a description of that structure is available. It is of even greater advantage if the structure is independent of any specific hardware or software and machine-processible. There is also a distinction between formats used for raw data, which are largely binary, and metadata and derived data, such as catalogues, which are more structured and predominantly textual.

Raw binary formats tend to be domain specific, although there is some usage of FITS outside of astronomy. In common with other formats, such as HDF5, descriptions of the binary structures and their metadata (often combined) are separable. CDF and netCDF  take the concept even further by defining a common data model for scientific data sets, which has its own associated API. This handles data reading, the coordinate systems the data are expressed in and specific types of data and divorces the data user entirely from the physical details of its storage.

The most familiar textual data representations are XML and JSON and systems exist to describe the structures of these, e.g., XML Schema and JSON Schema. A frequent criticism of them, however, is that they are ineffectual formats, particularly where processing speed is a factor since this is done on a character-by-character basis. Bandwidth and storage issues can also be significant and binary versions are not necessarily any better. Google's Protocol Buffers follows a similar abstraction path to CDF/netCDF and was designed to be a faster alternate to XML. Data structures are defined in terms of a common data model with an API for access and manipulation. The actual format of the underlying data is immaterial --  the default is binary, but textual formats may also be used --, the libraries provide the necessary interfaces to it. Apache Avro follows a similar approach, employing JSON to define its data structures but only using a compact binary data format.  

When it comes to communicating data, actual physical transportation -- the so-called sneakernet method -- remains one of the most efficient and reliable means, sacrificing latency for high throughput, and employed by many large astronomy projects as well as commercial service providers. However, every instance remains a bespoke solution, defying standardization, with the exact details only known between the sender and the receiver. When the desire is to communicate data to potentially millions anywhere and at any time, alternate solutions are required. 

Despite living at the time of greatest interconnectivity in human history, the existing infrastructure is insufficient for our needs: we've officially run out of IPv4 addresses and the Internet pipes are straining under the pressures of streaming media. Next generation efforts, such as Internet2, are developing the advanced capabilities that are required, e.g., the on-demand creation and scheduling of high-bandwidth, high-performance data circuits but the use of the current setup can also be optimized. 

Conventional data transfer technologies rely on a single stream/channel between the sender/provider and the receiver to carry the data, which typically does not make full use of the available bandwidth. Chunking up the data and sending it over multiple streams to the receiver achieves a much greater use of bandwidth, e.g., GridFTP works in this way. These streams can either come from multiple providers, each with their own (partial) copy of the data (just the requested chunk needs to be available), or a single provider running parallel streams. Chunks are requested from providers based on their advertised availability and, once the receiver has a chunk, it can also become a provider for it -- this is the basis for many peer-to-peer transport systems.

Data streams typically use TCP packets for their transport but this can exhibit poor performance in long distance links, particularly when the bandwidth is high, or when multiple concurrent flows are involved with different data transfer rates. UDT employs UDP packets instead to achieve much faster rates than TCP can but with its same reliability. Other solutions involve fine-tuning TCP specifically for high performance networks or modifying the TCP protocol.

Not all data formats encode their information in as efficient a manner as achievable and it is often possible to reduce the size of a data object for transmission (or storage) by compressing it. Significant improvements can be achieved, particularly for textual data, with generic compression routines such as gzip and bzip2. For astronomical binary data -- images and tables -- FITS tile compression \cite{fitscompression} offers better performance than these
and also preserves the FITS headers (structure description) uncompressed for faster access. In fact, with the appropriate library (CFITSIO), compressed data should be the default mode for operation with decompression never being necessary.

With larger amounts of data, storage and bandwidth become a premium -- OAIS (ISO 14721; 2003) is a high-level and well-regarded model for the complete archival cycle, useful for framing discussions about, and critiquing, data management planning. The meaning of the data, however,  lies in its inherent structure and making this independent of the actual arrangement of bytes is no different to abstracting the meaning of creation from its encoding in the world around us.

\section{The rhetoric of data} % How is it communicated
\label{sec:4}
Students in the Middle Ages were drilled by rote in the skills of writing letters and sermons, drawing on the rhetorical teachings of classical antiquity. It was presupposed that the structure of language corresponded to that of being and understanding and therefore the manner and style of communicating well and correctly was important, employing the appropriate tone and linguistic constructs for the given subject matter (an appreciation that contributed to the scientific method). Data is the language of the scientific dialectic and highly politicized with a suite of tricks and devices to 
lead an audience to a particular conclusion.

The credibility of an interpretation is as much a function of how it has been reached as it is a matter of trust in the data upon which it is based. The level of such trust, however, seems to be inversely proportional to how easy the data is to access. Though astronomical data is commercially valueless (described by Jim Gray as a zero billion dollar problem), most of it still resides in protected vertical silos, accessible for long periods of time to only an elect handful. Attempts to create an open data culture are viewed as seditious: the only contemporary survey to make its data publicly accessible from the outset is the Catalina Real-Time Transient Survey~\cite{crts}. 

This level of control persists even when the results have undergone peer review and appeared in the public domain. 
It can be a Sisyphean task to get supporting data to replicate or build upon particular interpretations. In the life sciences, generally all data is required to be made available without precondition when an associated paper is published, preferably in a community-endorsed public data repository. Astronomy already has a culture of data centres but these tend to be too tied to specific big missions or wavelength regimes -- there is certainly no current repository where arbitrary data can be archived or even permanently registered.

The glacial progress of traditional astronomical publishing is countered by online bibliographic services, such as arXiv and ADS, which provide access points to associated data where available. An even more recent trend is the pre-submission discussion of data on blogs and other social networking fora. Although much of this is clearly intended for the sake of publicity rather than serious scientific discourse, it does reflect a growing frustration with the existing peer review system, particularly in our increasingly connected and open society, and the interest in alternatives such as open peer review and open peer commentary.

The progressive emergence of interdisciplinary fields is also challenging since data is often taken out of its original context and reused in an entirely new (and, maybe, not entirely appropriate) one. This so-called pick-and-mix model allows one far greater latitude to present (apparently) supported conclusions, either intentionally or, more usually, by accident, in areas where there is a current lack of domain expertise. As mentioned previously, however, the formal use of semantics can go some way to preventing this.

For a thousand years, data has been a precious commodity, residing in select locations and to be safeguarded at all costs. The necessity of an open approach in the new era stands against the existing control and access structures but is far more in tune with the intended purity and selflessness of the scientific method. Data should be free to all.

\section{The arithmetic of data} % Data in its own right (pure)
\label{sec:5}
From the abacus to the algorithm, arithmetic was concerned less with reckoning than with understanding the nature of number, its properties, and the uniqueness of numerical series obtained by certain constant relationships. It was far more qualitative than quantitative, motivated by a desire to divine the presence of an unseen hand in Nature expressed in the beauty of Platonic perfection. Whilst we do not seek transcendence in data, exploring its nature and its properties is still an illuminating experience.

The utility (or value) of data lies in its ability to convey information (although one person's data can be another person's noise). This is a highly variable quantity, dependent on the size and potential impact of its contents, i.e., how supportive or challenging they are to the current paradigm, as well as its timeliness. The relative utility of individual pieces of data can be ranked, producing an overall trend that is logistic: initial data in an area is approximately exponential in utility, e.g., observations of 10 Type Ia supernovae (SNe Ia) in the redshift range $0.16 \le z \le 0.62$ suggest an accelerating universe \cite{riess}; then, as progressively more data becomes available, saturation occurs and its utility slows, e.g., successive observations supporting the SNe Ia results; and at maturity, it has essentially zero utility, e.g., surveys are regularly showing consistent behaviour. The metatrend may well be a succession of logistic behaviours or approaching something that is multiply logistic, depending on how much new paradigms redefine the utility of old data. 

Unprecedented progress along these logistic trends is being driven by two factors. Firstly, the future is characterized by massive parallel streams of (small) data events rather than large monolithic slabs of data. The synergistic nature of data (as expressed in Szalay's law that the utility of $N$ comparable data sets is $N^{2}$) means that these streams  clearly lead to potentially rapid progress along the logistic curve, providing that they are linkable. Paradoxically the advent of the data-intensive era marks the inflection point in utility growth for single data sets. 

Secondly, there is the increasing pace of data acquisition, driven by exponential growth rates in technology (in particular, Moore's law regarding the transistor density of integrated circuits). Some believe that these rates cannot continue indefinitely: at some stage, either the relative doubling times of different technologies will become incompatible -- the slowest one defining the breaking point --, or one of them will come up against the hard edge of some physical law, or the economics of continued growth will cease to be attractive or advantageous. Others feel that new technologies will arise to keep the exponential growth up at equivalent rates, at least, if not accelerating ones.

Power considerations are an increasingly important aspect. Already in 2004, microprocessor clock rates flatlined owing to power dissipation limits, although increasing the number of cores per chip has maintained the growth rate for computational performance. Exascale systems (desktop petaflop/embedded teraflop) have predicted power needs of $\sim$100 MW \cite{exascale} but even commodity-level processors are heading towards a power wall. One mitigating strategy is to employ GPUs for as much general purpose computation as possible \cite{gpu} -- they offer far better flop/Watt performance than CPUs. However, they must be supported by a CPU to run the operating system and manage the GPU device. Using a low-power CPU processor, which would spend much of its time idling, is a viable short-term solution but, inevitably, trans-silicon technologies will need to be considered -- these require lower energy but at a cost of slower clock speeds.

If the universe is fundamentally reducible to a set of equations then there is a finite amount of information to be extracted from data. The extent to which we can approach that limit is determined by the technology and energy available to us in that pursuit, although ultimately the law of diminishing returns may still render it unattainable. 
If, however, the world is unknowable then gathering data and deriving information from it are endless activities.

\section{The geometry of data} % Data in space
\label{sec:6}
The great cathedrals of mediaeval Europe were intended as sacred mirrors of creation, reflecting the design and structure of the universe through the laws and forms of geometry, translated by the master stonemason in imitation of the work of his divine master. By the same token, the great data centers of tomorrow will reflect the aspirations of master scientists and technologists to facilitate the study of the design and structure of the universe through the laws and forms of a new geometry, the architectural order of vast collections of data. 

The physical media of sacred geometries are well understood, be it Caen stone and Purbeck marble or hard drives. 
Petascale storage systems can be constructed from commodity Terabyte-sized components for approximately $\$$50000/PB at the time of writing, although suitable precautions must be taken to protect against the high failure rates and subsequent data loss that are associated with "cheap" commodity disks. The art and skill then lies in layering the data on these in as efficient and effectual a manner as possible according to user constraints. 

A standard architecture for high throughput data that is intended to be predominantly read and rarely overwritten or appended (e.g., for data processing) is to break it up into fixed size chunks (typically 64 MB) and then distribute multiple copies (typically three, two on the same rack and one on a different one)  of each chunk across the disk cluster (see, for example, Google FS \cite{gfs} or its open-source equivalent, HDFS). This provides reliability against the potential inadequacies of the underlying hardware and can be fine-tuned (more copies) for specific data where greater demand or protection is anticipated. A central/master node maintains the list of which chunk is where and any attendant metadata, and also the list of all operations involving data. This does, however, present an obvious single point of failure and can limit scalability (distributing the master node is a possible solution).

Such systems are optimized for very large data sets with a small number of constituent parts.  When there are large numbers of small files in a data set, the dominant process during runtime execution of a computation on that data set  is locating the relevant chunks, i.e., calls to the master node \cite{wiley}. HDFS mitigates this by defining a specific data structure for such situations -- the sequence file, which is essentially a container of smaller files bundled with an index -- vastly reducing the number of files on disk that need to be processed. Further improvements can be obtained by structuring sequence files according to some prescription, e.g., spatial or temporal location of image files, rather than just randomly grouping files into them. 

Alternate data scenarios involve low-latency random access (high availability) to the data, e.g., retrieving thumbnail images, or very large numbers of varying sized files with multiple concurrent writes, e.g., log files. In these cases, approaches based around distributed multi-dimensional sorted maps, such as Google's BigTable \cite{bigtable} or Hadoop's open-source equivalent, HBase (both built on top of GFS and HDFS respectively), or more general distributed data and metadata architectures, such as OpenStack Swift or iRODS, are more  appropriate. 
			
All these physical architectures broadly have no knowledge of the structure of the data that they are dealing with. However, there is a subclass that is concerned specifically with the type of data that one would traditionally put in a (relational) database (RDBMS). RDBMSs do not function well beyond $\sim$100 TB in size \cite{gray} but there is a clear need for equivalent systems to support petascale catalogs, etc. BigTable and its variants belong to a superclass of systems known as NoSQL, which provide distributed storage for structured data, and can used as scaled equivalents to databases for many types of data. However, a better match for scientific data is afforded by SciDB which is a column-oriented (rather than row-oriented like a RDBMS) system that uses arrays as first-class objects rather than tables and is still ACID (like a RDBMS but unlike most NoSQL solutions).

The intricate geometries that we employ in our data centers with replicated hierarchical patterns are no different to those used by stoneworkers ten centuries ago in their own towering edifices. Both are intended to reflect our knowledge of the design and structure of the universe itself, expressed in human works.

\section{The music of data} % Data in time
\label{sec:7}    
The ancients believed that the heavens were pervaded by the harmony of the spheres, the majestic fugue created by the movements of the celestial bodies. The mediaeval curriculum formalized this, along with the internal fugue of the human body and the audible fugues that we could create, into the concept of {\em musica}, which studied the progression of proportions through time according to well-established patterns and rules. The progression of data through time as a result of computations on it is a similar fugue and, in the case of large data sets, there are a number of identifiable patterns.

The predominant such computational pattern today is the so-called embarrassingly parallel task, which describes a computation for which little or no effort is required to separate it into a number of parallel tasks, often with no dependency between them, e.g., anything requiring a sweep through a parameter space. These can then be distributed across the available processors, bringing a substantial reduction to the computation time in comparison with a straightforward sequential approach. If the processors can be selected so that the data they require is local (data locality), this further reduces the computation time (in fact, this is a general principle with large data sets -- bring the computation to the data). 

Several frameworks exist for managing these computations: Condor and BOINC will handle generic jobs on general pools of machines, ranging from local resources dedicated to the process to spare cycles scavenged from online resources anywhere in the world (the usual scenario for BOINC), although data is invariably transferred to the computation with these. Note that GPUs offer an increasingly popular alternative to CPU clusters
with single high-end chips offering performance speed-ups of up to $\sim$1000 compared to single CPUs, assuming appropriate code parallelization. In fact, GPU clusters make bulk brute force calculations viable over state-of-the-art CPU algorithmic approaches, for example, in $n$-point correlation functions \cite{tian}.

MapReduce \cite{mapreduce}, and its open-source equivalent, Hadoop, take a different approach by expressing jobs in terms of two standard operations -- map and reduce, instances of which (mappers and reducers) are deployed to the compute resources holding the data to be processed (thus ensuring data locality). A mapper
transforms its input data (as (key, value) pairs) to an intermediate set of different (key, value) pairs. Gathering these from all mappers, they are reordered and the group of data for each different key is sent to a reducer. Finally the outputs of the reducers are collected and returned as the overall result of the computation.

Not all computations are expressible in this form --  those which require a large amount of state information to be shared between mappers, e.g., referencing a common training set, with a lot of fine-grained synchronization can be problematic, although those involving iterative processes can often be expressed as chains of MapReduce tasks. An alternate pattern is to apply a streaming solution to the computation, i.e., one which only requires a single pass through the data. Typically these involve an incremental (online) formulation of the computational algorithm which updates with each new data point. Further optimizations are possible for specific types of computation, such as stochastic gradient descent for some types of machine learning. Obviously for large data sets, computations based on a single reading of the data are ideal and, in some cases, such algorithms also lend themselves to parallelization.

In the same way that polyphony lay at the heart of the mediaeval fugue with multiple voices combining to form a harmonic whole, parallelization is at the core of the modern data fugue with thousands of cores and threads acting in concert to transform vast data sets into harmonic representations of our knowledge of the cosmos.

\section{The astrology of data} % Data in space and time
\label{sec:8}
``As above, so below'' underpinned the mediaeval conviction that patterns in the heavens reflected, or even presaged, happenings here on Earth in all spheres of life, from personal health to affairs of state to triumphs and disasters. {\em Astronomia} was both the science of observing these patterns and interpreting them, drawing on the corpora of Babylonian and Islamic thought. The plans for creation were writ large in the celestial arrangements of stars and planets and we could divine them by proper study. Data mining is ``the semi-automatic discovery of patterns, associations, changes, anomalies, and statistically significant structures and events in data'' \cite{kddig} and is the mainstay of astroinformatics.

The application of data mining to a data set really has two primary goals \cite{fayyad}: predicting the future behaviour of certain entities based on the existing behaviour of other entities in the data (prediction) and finding human-interpretable patterns describing the data (description) -- interestingly the same division distinguished judicial and natural astrology. The suite of available data mining techniques, originating primarily from computer science (particularly artificial intelligence research) and statistics, can then be regarded as falling into one or more of these categories: classification, regression, clustering, summarization, dependency modelling, and change and deviation (or outlier) detection.

The process of data mining extends well beyond just the casual employment of a particular algorithm, however. The data of interest first has to be collected and carefully prepared for analysis, e.g., normalization, handling missing values, binning, sampling, etc. The assumptions and limitations of the particular technique that is going to be applied have to be assessed, e.g., the specific number of clusters to be defined, and, in many cases, this will require multiple applications of the algorithm to fully determine these. Even then, the outcome has to be validated, either by rerunning the analysis on subsets of the data and/or using some particular measure of quality. Finally, the procedure is understood well enough that results can be interpreted and it can be used with further and wider data samples.

An important aspect of data mining is the incorporation of appropriate prior knowledge. Statistical inferencing (see section \ref{sec:2}) is one approach to this but it builds its arguments on probabilistic models of the data and not on the actual observed values. Thus its interpretations rest on the assumption that the model is a good description of reality and not on the observations. Folding the knowledge into the data mining algorithm at least means that any interpretations are data-based, even if the knowledge might be model-derived. From semantic constructs, such as ontologies, similarity metrics can be defined which encode the degree to which two concepts share information. These quantitative measures of conceptual similarity can be then be incorporated into standard data mining algorithm formulations, giving knowledge-driven data mining.

Of all the patterns discerned in the heavens by mediaeval scholars, the most vital was the {\em computus}, which allowed the determination of the date of Easter. The utility of the patterns that we have discovered in astronomical data has led to the discovery of new objects, improved processing, object detection and classification, and better photometric redshifts \cite{kddguide}.

\section{The scholasticism of data}
\label{sec:9}
The {\em trivium} and the {\em quadrivium} created a scholastic culture in which all phenomena, both natural and artificial, were subject to interrogation and symbolic interpretation. The liberal arts not only conferred the necessary skills to uncover the knowledge hidden throughout creation but provided a framework onto which these discoveries could be attached and understood. In particular, the properties and relationships of numbers, unchanging and endless, were a path to divine revelation. Our desire to reveal the inner workings of the universe is unchanged but we no longer require it to be numinous. The scientific method which arose out of the dialectic criticisms of the Middle Ages is founded on rational thought and logic, dealing with hard data and facts, rather than association and exegetical consistency.

We have shown, however, how the same themes run through our contemporary approach. In our vast data sets, 
we are still concerned with the structures that we employ to represent our knowledge, communicating them well and correctly, and how we can meaningfully design and make them. We still need to understand what it is that we are studying and what rules apply. And we still need to know how to look for the meaningful patterns that we want to uncover. Only with this grounding can we hope to manage the overwhelming volumes and complexities of data that are facing us. 

Finally, this has to be a community effort, both international and interdisciplinary. The challenges for astronomy are the same for climate science, for genomics, for any 21$^{st}$ century enterprise. Efforts such as the International Virtual Observatory Alliance \cite{ivoa} are a step in the right direction but we need something that is truly universal, educating at all levels and in all subjects. Data, like its mediaeval counterpart, number, must be a first-class entity in our worldview, and not just from a technological standpoint. From a future vantage point, today will be regarded as the point from which we emerged from the Dark Ages of data and initiated a truly modern perspective.

\subsubsection*{Acknowledgments}
We would like to thank Norman Gray and Helen Angove for useful feedback and discussion about this paper.


\begin{thebibliography}{99.}%
% Use the following syntax and markup for your references if 
% the subject of your book is from the field 
% "Mathematics, Physics, Statistics, Computer Science"
%
\bibitem{fp} Hey, T., Tansley, S., Tolle, K.: The Fourth Paradigm. Microsoft Research, Redmond (2009) 
%
\bibitem{kline} Kline, M.: Mathematics in Western Culture. Oxford University Press, Oxford (1953)
%
\bibitem{norman} Gray, N.: Data is a singular noun (2010);
\url{http://purl.org/nxg/note/singular-data.}
%
\bibitem{hogglang1} Hogg, D.~W., \& Lang, D.: Astronomical imaging: The theory of everything. American Institute of Physics Conference Series \textbf{1082}, 331 (2008) 
%
\bibitem{hogglang2} Hogg, D.~W., \& Lang, D.: Telescopes don't make catalogs! EAS Publications Series \textbf{45} 351 (2011)
%
\bibitem{fitscompression} FITS image compression programs; \url{http://heasarc.gsfc.nasa.gov/fitsio/fpack}
%
\bibitem{crts} Drake, A. et al.: First Results from the Catalina Real-Time Transient Survey. ApJ \textbf{696}, 870 (2009)
%
\bibitem{riess} Riess, A. et al.: Observational Evidence from Supernovae for an Accelerating Universe and a Cosmological Constant . AJ \textbf{116}, 1009 (1998)
%
\bibitem{exascale} Kogge, P. et al.: ExaScale Computing Study: Technology Challenges in Achieving Exascale Systems. (2008) doi: 10.1.1.165.6676
%
\bibitem{gpu} Fluke, C.J., Barnes, D.G., Barsdell, B.R., Hassan, A.H.:Astrophysical Supercomputing with GPUs: Critical Decisions for Early Adopters. PASA \textbf{28}, 15--27 (2011)
%
\bibitem{gfs} Ghemaway, S., Gobioff, H., Leung, S.T.: The Google File System (2003);
\url{http://labs.google.com/papers/gfs.html}
%
\bibitem{wiley} Wiley, K., Connolly, A., Gardner, J., Krughoff, S., Balazinska, M., Howe, B., Kwon, Y., Bu, Y.: Astronomy in the Cloud: Using MapReduce for Image Co-Addition. PASP \textbf{123}, 366--380 (2011)
%
\bibitem{bigtable} Chang, F., et al.: Bigtable: A Distributed Storage System for Structured Data (2006); \url{http://labs.google.com/papers/bigtable.html}
%
\bibitem{gray} Gray, J., Hey, T.: In Search of PetaByte Databases (2001);\\
\url{http://www.research.microsoft.com/~Gray/talks/In\%20Search\%20of\%20PetaByte\%20Databases.ppt} 
%
\bibitem{tian} Tian, H.J., Neyrinck, M.C., Budavari, T., Szalay, A.S.: Redshift-space Enhancement of Line-of-sight Baryon Acoustic Oscillations in the Sloan Digital Sky Survey Main-galaxy Sample. ApJ \textbf{728}, 34 (2011)
%
\bibitem{mapreduce} Dean, J., Ghemawat, S.: MapReduce: Simplified Data Processing on Large Clusters (2004); \url{http://labs.google.com/papers/mapreduce.html}
%
\bibitem{kddig} IVOA Knowledge Discovery in Databases; \\
\url{http://www.ivoa.net/cgi-bin/twiki/bin/view/IVOA/IvoaKDD}
%
\bibitem{fayyad} Fayyad, U., Piatetsky-Shapiro, G., Smyth, P.: From Data Mining to Knowledge Discovery in Databases. AI Magazine \textbf{17}, 37--54 (1996)
%
\bibitem{kddguide} Data mining examples; 
\url{http://www.ivoa.net/cgi-bin/twiki/bin/view/IVOA/IvoaKDDguideScience}
%
\bibitem{ivoa} International Virtual Observatory Alliance (IVOA); \url{http://www.ivoa.net}


% Contribution 
%\bibitem{science-contrib} Broy, M.: Software engineering --- from auxiliary to key technologies. In: Broy, M., Dener, E. (eds.) Software Pioneers, pp. 10-13. Springer, Heidelberg (2002)
%
% Online Document
%\bibitem{science-online} Dod, J.: Effective substances. In: The Dictionary of Substances and Their Effects. Royal Society of Chemistry (1999) Available via DIALOG. \\
%\url{http://www.rsc.org/dose/title of subordinate document. Cited 15 Jan 1999}
%
% Monograph
%\bibitem{science-mono} Geddes, K.O., Czapor, S.R., Labahn, G.: Algorithms for Computer Algebra. Kluwer, Boston (1992) 
%
% Journal article
%\bibitem{science-journal} Hamburger, C.: Quasimonotonicity, regularity and duality for nonlinear systems of partial differential equations. Ann. Mat. Pura. Appl. \textbf{169}, 321--354 (1995)
%
% Journal article by DOI
%\bibitem{science-DOI} Slifka, M.K., Whitton, J.L.: Clinical implications of dysregulated cytokine production. J. Mol. Med. (2000) doi: 10.1007/s001090000086 
%

\end{thebibliography}
\end{document}